\documentclass[12pt,preprint]{elsarticle}
\usepackage{amsmath}
\usepackage{amssymb}

\usepackage{color}
\newtheorem{theorem}{Theorem}[section]

\newtheorem{remark}[theorem]{Remark}\numberwithin{equation}{section}

\journal{}

\begin{document}

\begin{frontmatter}

\title{Group analysis approach for finding reciprocal transformations for the two-dimensional stationary gasdynamics}

\author[MU]{Piyanuch Siriwat}

\ead{piyanuch.sir@mfu.ac.th}

\author[SUT]{S.V.~Meleshko \corref{cor1}}

\cortext[cor1]{Corresponding author}

\ead{sergey@math.sut.ac.th}

\address[SUT]{School of Mathematics, Institute of Science, Suranaree University
of Technology, 30000, Thailand}

\address[MU]{School of Science, Mae Fah Luang University, Chiang Rai, 57100,
Thailand}
\begin{abstract}
Equivalence transformations play one of the important roles in continuum
mechanics. These transformations reduce the original equations to
simpler forms. One of the classes of nonlocal equivalence transformations
is the class of reciprocal transformations. Despite the long history
of applications of such transformations in continuum mechanics, there
is no method of obtaining them. Recently such a method was proposed
by the second author of the present paper. The method uses group analysis
approach and it consists of similar steps as for finding an equivalence
group of transformations. The new method provides a systematic tool
for finding classes of reciprocal transformations (group of reciprocal
transformations). As an illustration, the method was applied to the
one-dimensional gas dynamics equations, and new reciprocal transformations
were found. Similar to the classical group analysis this approach
can be also applied for finding all reciprocal transformations (not
only composing a group) of studied equations. The present paper provides
this algorithm. As an illustration the method is applied to the two-dimensional
stationary gas dynamics equations. Equivalence group, reciprocal equivalence
group and completeness of all discrete reciprocal transformations
are presented in the paper.
\end{abstract}
\begin{keyword}
Reciprocal transformations, equivalence group of transformations,
Lie group of transformations

Subject Classification (MSC 2010): 35C99, 76W05

\end{keyword}
\end{frontmatter}

\section{Introduction}

In a study of lift and drag aspects in two-dimensional homentropic
irrotational gasdynamics Bateman \cite{art:Bateman1938} established
invariance of the governing system under a novel multi-parameter class
of relations which have come to be known as reciprocal transformations.
The latter are typically associated with conservation laws admitted
by a system. These transformations leave invariant the governing equations,
up to the equation of state. In the group analysis method such kind
of transformations are called equivalence transformations. In nonlinear
continuum mechanics, reciprocal transformations have likewise proved
to have diverse physical applications\footnote{See, for example in the books \cite{bk:RogersShadwick1982,bk:MeirmanovPukhnachovShmarev1997}.}.
The preceding attests to the importance of reciprocal transformations
in physical applications. Despite on many applications, to the best
of our knowledge, there are no systematic derivation of reciprocal
transformations.

A link between a one-parameter subclass of infinitesimal reciprocal-type
transformations in gasdynamics and the Lie group approach was established
in \cite{art:IbragimovRogers2012}, where it was shown that the transformations
found in \cite{art:Rogers1968} can compose a Lie group of transformations.
This idea was also applied in \cite{art:MeleshkoRogers2021} for
relativistic gas dynamics equations, where a connection between one-parameter
subclasses found recently in \cite{art:RogersRuggeri2020,art:RogersRuggeriSchief2020}
and a Lie group procedure was shown.

\subsection{Equivalence transformations}

A nondegenerate change of the dependent and independent variables,
which transfers any system of differential equation of the given class
\begin{equation}
F^{k}(x,u,p,\phi)=0,(k=1,2,\ldots,s)\label{eq:sec5_2.1}
\end{equation}
to the system of equations of the same class is called an equivalence
transformation. Here $x$ is the vector of the independent variables,
$u$ is the vector of the dependent variables, $p$ are derivatives,
and the functions $\phi:V\to R^{t}$ are arbitrary elements of system
(\ref{eq:sec5_2.1}). For example, for the gas dynamics equations
$\phi$ is the function related with the state equation.

\subsubsection{Equivalence point transformations}

The problem of finding an equivalence transformation consists of the
construction a transformation of the variables $(x,u,\phi)$ that
preserves the equations changing only their representative $\phi=\phi(x,u)$.
For this purpose there are several methods. One of these methods is
the direct solution of the equations determining such transformations
(see for example \cite{art:KingstonSophocleous1998}). Despite its
complexity this method gives a complete set of the equivalence point
transformations \cite{art:VaneevaJohnpillaiPopovychSophocleous}.
The determining equations become simpler for the equivalence transformations
composing a Lie group \cite{bk:Ovsiannikov1978}, which is called
an equivalence group. Notice that using equivalence group and the
method proposed in \cite{art:Hydon1998,art:Hydon2000}, all equivalence
transformations can be found \cite{bk:BihloBihloPopovych2015}\footnote{See also references therein.}.

\subsubsection{Reciprocal transformations}

For reciprocal transformations the change of the independent variables
differs from point transformations: this change is defined by differentials.
A link between a one-parameter subclass of infinitesimal reciprocal-type
transformations in gasdynamics and the Lie group approach was established
in \cite{art:IbragimovRogers2012}. The results of \citep{art:IbragimovRogers2012}
led to the idea to use the algorithms developed in the group analysis
method \cite{bk:Ovsiannikov1978} for finding reciprocal transformations.
This idea was realised in \cite{art:Meleshko2021}, where reciprocal
transformations and equivalence transformations of the one-dimensional
gas dynamics equations were compared. The reciprocal transformations
obtained by this way we call group of reciprocal transformations.

In the present paper, for finding all reciprocal transformations of
systems of differential equations we combined the method of finding
reciprocal group of transformations \cite{art:Meleshko2021} with
the automorphism-based algebraic method \cite{art:Hydon1998,art:Hydon2000}\footnote{See also \cite{art:CardosoPopovych2013,art:KontogiorgisPopovychSophocleous,bk:BihloPoltavetsPopovych2020}
for further development and extensions of the method.}. This combination allows finding all reciprocal transformations (not
only composing a group). For an illustration the method is applied
to the two-dimensional stationary gas dynamics equations. Equivalence
group, group of reciprocal transformations, and all discrete reciprocal
transformations are presented in the paper. It is proven that the
reciprocal transformations found in \citep{art:PowerSmith1961} compose
a complete set of reciprocal transformations, up to equivalence transformations.

\subsection{The two-dimensional gas dynamics equations }

The equations are considered in the present paper are \citep{bk:Ovsiannikov[2003]}
\begin{equation}
\begin{array}{c}
F_{1}=(\rho u)_{x}+(\rho v)_{y}=0,\,\,\,F_{2}=\rho(uu_{x}+vu_{y})+p_{x}=0,\\
F_{3}=\rho(uv_{x}+vv_{y})+p_{y}=0,\,\,\,F_{4}=uS_{x}+vS_{y}=0,
\end{array}\label{eq:2Dgas}
\end{equation}
where $\boldsymbol{q}=(u,v)$ is the gas velocity, while $p$ is the
gas pressure, $\rho$ is the gas density and $S$ is the specific
entropy. An appropriate state equation must be added to system (\ref{eq:2Dgas})
\begin{equation}
{\displaystyle p=G(\rho,S)}.\label{eq:2.2}
\end{equation}

\subsection{Organization of the paper}

The paper is organized as follows.

The next Section deals with the equivalence group of the two-dimensional
stationary gas dynamics equations. In Section 3, the well-known results
related with the reciprocal transformations of the two-dimensional
stationary gas dynamics equations are presented. Section 4 discusses
applications of the first method for seeking reciprocal transformations
using the infinitesimal approach. This method uses two conservation
laws written in the form of differentials. Section 5 provides a generalization
of the first method, where none of the assumptions on the differentials
are required. Section 6 is devoted to the application of the results
of the previous section for finding all reciprocal transformations
for the two-dimensional stationary gas dynamics equations. Summary
of the results is also given in this Section. The last Section gives
the Conclusions. Some necessary formulas are given in the Appendix.


\section{Equivalence point transformations}

Equivalence transformations preserve the structure of equations. Consider
an equivalence group preserving system (\ref{eq:2Dgas}). A generator
of a one-parameter group of equivalence transformations is assumed
to be in the form \cite{bk:Ovsiannikov1978,bk:Meleshko[2005]}
\[
X^{{\rm e}}=\xi^{x}\partial_{x}+\xi^{y}\partial_{y}+\zeta^{\rho}\partial_{\rho}+\zeta^{u}\partial_{u}+\zeta^{v}\partial_{v}+\zeta^{S}\partial_{S}+\zeta^{p}\partial_{p},
\]
where all coefficients of the generator depend on $(x,y,\rho,u,v,S,p)$,
and ${\displaystyle p=G(\rho,S)}$ is considered as an arbitrary element.

For finding the equivalence group of transformations the infinitesimal
criterion \cite{bk:Ovsiannikov1978} is used. For this purpose the
generator $X^{e}$ is prolonged, and using the prolonged generator,
applied to equations (\ref{eq:2Dgas}), the determining equations
of the equivalence transformations are derived. The solution of these
determining equations gives the general form of the elements of the
equivalence group. Calculations show that the basis elements of the
corresponding Lie algebra are
\begin{equation}
\begin{array}{c}
X_{1}^{e}=\partial_{x},\,\,\ X_{2}^{e}=\partial_{y},\,\,\,X_{3}^{e}=-v\partial_{u}+u\partial_{v}-y\partial_{x}+x\partial_{y},\,\,\,X_{4}^{e}=x\partial_{x}+y\partial_{y},\\
X_{5}^{e}=\rho\partial_{\rho}+p\partial_{p},\,\,\,X_{6}^{e}=\partial_{p},\,\,\,X_{h}^{e}=h(S)(-2\rho\partial_{\rho}+u\partial_{u}+v\partial_{v}),\,\,\,X_{F}^{e}=F(S)\partial_{S}
\end{array}\label{eq:may06-1}
\end{equation}
where $h(S)$ and $F(S)$ are arbitrary functions.

The transformations corresponding to the generators $X_{i}^{e}$,
($i=1,2,...,4)$ and $X_{F}$ are well-known in the gas dynamics.
The generators $X_{1}^{e}$, and $X_{2}^{e}$ define the shifts with
respect to $x$ and $y$, respectively, $X_{3}^{e}$ corresponds to
the rotation transformation, $X_{4}^{e}$ and $X_{5}^{e}$ define
scalings, $X_{6}^{e}$ corresponds to the shift of $p$. The generator
$X_{h}^{e}$ defines the projective transformations
\begin{equation}
\rho^{\prime}=\rho\psi^{-2},\,\,\,u^{\prime}=u\psi,\,\,\,v^{\prime}=v\psi,\label{eq:aug02.1}
\end{equation}
where $\psi=e^{\epsilon h}$, $\epsilon$ is the group-parameter,
and only changeable variables are presented. This transformation corresponds
to the Munk\textendash Prim transformation \citep{art:MunkPrim1947},
which is also well-known in the gas dynamics. The transformations
related to the generator $X_{F}^{e}$ allow one to change $S^{\prime}=\Psi(S)$,
where $\Psi(S)$ is an arbitrary function.

There are also known two obvious involutions
\[
E_{1}:\,\,\,\,x^{\prime}=-x,\,\,\,u^{\prime}=-u,
\]
\[
E_{2}:\,\,\,\,y^{\prime}=-y,\,\,\,v^{\prime}=-v.
\]

\section{The Bateman-type reciprocal transformations of 2D stationary gasdynamics}

System (\ref{eq:2Dgas}) implies the pair of conservation laws
\begin{equation}
\begin{array}{c}
(\rho uv)_{x}+(p+\rho v^{2})_{y}=0,\\
(p+\rho u^{2})_{x}+(\rho uv)_{y}=0.
\end{array}\label{eq:2.3}
\end{equation}
Using these conservation laws, new independent variables can be introduced
by the formulas
\begin{equation}
\begin{array}{c}
dx^{\prime}=\beta_{1}^{-1}[(p+\beta_{2}+\rho v^{2})dx-\rho uvdy],\\
dy^{\prime}=\beta_{1}^{-1}[-\rho uvdx+(p+\beta_{2}+\rho u^{2})dy]
\end{array}\label{eq:2.4}
\end{equation}
subject to the requirement that
\begin{equation}
\begin{array}{c}
p+\beta_{2}+\rho q^{2}\neq0.\end{array}\label{eq:2.5}
\end{equation}
In \citep{art:PowerSmith1961} it was established that the gasdynamic
system (\ref{eq:2Dgas}) is invariant under the 4-parameter class
of the Bateman-type reciprocal transformations
\begin{equation}
\begin{array}{c}
{\displaystyle u^{\prime}=\frac{\beta_{1}u}{p+\beta_{2}},\,\,\ v^{\prime}=\frac{\beta_{1}v}{p+\beta_{2}},}\\
{\displaystyle p^{\prime}=\beta_{4}-\frac{\beta_{1}^{2}\beta_{3}}{p+\beta_{2}},\,\,\ \rho^{\prime}=\frac{\beta_{3}\rho(p+\beta_{2})}{p+\beta_{2}+\rho q^{2}},\,\,\ S^{\prime}=F(S).}
\end{array}\label{eq:2.6}
\end{equation}
This result has its roots in work of Bateman on lift and drag functions
in planar irrotational gasdynamics \citep{art:Bateman1938}\footnote{Short review of the analysis and applications of these transformations
can be found in \citep{art:IbragimovRogers2012}. }.

In \cite{art:IbragimovRogers2012}, a link between a one-parameter
subclass of infinitesimal reciprocal-type transformations in gasdynamics
and a Lie group approach was established. It was observed that if
one sets
\begin{equation}
\begin{array}{c}
\beta_{1}=\beta_{2}=\beta_{4}=\epsilon^{-1},\,\,\ \beta_{3}=1,\end{array}\label{eq:2.7}
\end{equation}
then the one-parameter class of reciprocal transformations
\begin{equation}
\begin{array}{c}
{\displaystyle u^{\prime}=\frac{u}{1+\epsilon p},\,\,\ v^{\prime}=\frac{v}{1+\epsilon p},}\\
{\displaystyle p^{\prime}=\frac{p}{1+\epsilon p},\,\,\ \rho^{\prime}=\frac{\rho(1+\epsilon p)}{1+\epsilon(p+\rho q^{2})},\,\,\ S^{\prime}=F(S)}
\end{array}\label{eq:2.8}
\end{equation}
together with
\begin{equation}
\begin{array}{c}
dx^{\prime}=\epsilon[(p+\rho v^{2})dx-\rho uvdy]+dx,\\
dy^{\prime}=\epsilon[-\rho uvdx+(p+\rho u^{2})dy]+dy
\end{array}\label{eq:2.9}
\end{equation}
composes transformations similar to the Lie group of transformations.
This observation led to establishing a method \citep{art:MeleshkoRogers2021,art:Meleshko2021}
for constructing reciprocal transformations by using procedures developed
in the group analysis method. According to \citep{art:Meleshko2021},
transformations (\ref{eq:2.8}), (\ref{eq:2.9}) are called by the
group of reciprocal transformations. Similar to the classical theory
of a Lie group of point transformations, it is convenient to present
these transformations by the infinitesimal generators $X=F(S)\partial_{S}$
and
\[
Y=\rho^{2}q^{2}\partial_{\rho}+pu\partial_{u}+pv\partial_{v}+p^{2}\partial_{p}+\left(-(p+\rho v^{2})dx+\rho uvdy\right)\partial_{dx}+\left(\rho uvdx-(p+\rho u^{2})dy\right)\partial_{dy},
\]
where $F(S)$ is an arbitrary function.

For further presentation it is convenient to introduce the following
generators

\begin{equation}
\begin{array}{c}
X_{1}=-v\partial_{u}+u\partial_{v}-dy\partial_{dx}+dx\partial_{dy},\,\,\,X_{2}=dx\partial_{dx}+dy\partial_{dy},\\[2ex]
X_{3}=\rho^{2}q^{2}\partial_{\rho}+pu\partial_{u}+pv\partial_{v}+p^{2}\partial_{p}+\left(-(p+\rho v^{2})dx+\rho uvdy\right)\partial_{dx}\\
+\left(\rho uvdx-(p+\rho u^{2})dy\right)\partial_{dy},\\[2ex]
X_{4}=\frac{1}{2}\left(2p\partial_{p}+u\partial_{u}+v\partial_{v}-dx\partial_{dx}-dy\partial_{dy}\right),\,\,\,X_{5}=\partial_{p}.
\end{array}\label{eq:aug01.10}
\end{equation}

One notices that among these generators only the generator $X_{3}=Y$
substantially defines reciprocal transformations: the transformations
corresponding to other generators are equivalent to the transformations
belonging to equivalence group.

\section{The first infinitesimal approach} 

In \cite{art:Meleshko2021_1}, two methods are given for constructing
a group of reciprocal transformations.

The first method consists of two steps. First, using invariance of
the conservation laws, one finds the coefficients of the generator
of the group of reciprocal transformations related with the differentials.
Then, using this coefficients, one defines the prolongation formulas
for the group of reciprocal transformations. Applying the prolonged
generator to the studied differential equations, one derives determining
equations for the coefficients of the infinitesimal generator related
with the dependent and independent variables. The general solution
of the determining equations gives the set of generators of one-parameter
groups of reciprocal transformations. Solving the Lie equations corresponding
to these generators, one finds the one-parameter reciprocal transformations.
As in the classical group analysis method, the multi-parameter transformations
are constructed by the composition of the one-parameter reciprocal
transformations.


Equations (\ref{eq:2.3}) can be rewritten in the form of differentials
\begin{equation}
\begin{array}{c}
S_{1}=q_{11}\left((p+q_{12}+\rho v^{2})dx-(\rho uv+q_{13})\,dy\right)=0,\\
S_{2}=q_{21}\left(-(\rho uv+q_{23})\,dx+(p+q_{22}+\rho u^{2})dy\right)=0.
\end{array}\label{eq:Mar24.2}
\end{equation}
where $q_{ij},\,\,(i=1,2;\,\,j=1,2,3)$ are constant such that $q_{11}q_{21}\neq0$.

Consider the transformations defined by the generator
\begin{equation}
X=\zeta^{\rho}\partial_{\rho}+\zeta^{u}\partial_{u}+\zeta^{v}\partial_{v}+\zeta^{p}\partial_{p}+\zeta^{S}\partial_{S}+\zeta^{dx}\partial_{dx}+\zeta^{dy}\partial_{dy},\label{eq:apr11.5-1-1}
\end{equation}
where the coefficients $\zeta^{\rho}$, $\zeta^{u}$, $\zeta^{v}$,
$\zeta^{p}$ and $\zeta^{S}$ depend on the variables $(x,y,\rho,u,v,p,S)$,
$\zeta^{dx}$ and $\zeta^{dy}$ are linear forms with respect to the
differentials $dx$ and $dy$ with the coefficients $^{x}\zeta^{dx}$,
$^{y}\zeta^{dx}$, $^{x}\zeta^{dy}$ and $^{y}\zeta^{dy}$ also depending
on the variables $(x,y,\rho,u,v,p,S)$:
\[
\zeta^{dx}={}^{x}\zeta^{dx}\,dx+{}^{y}\zeta^{dx}\,dy,\,\,\,\zeta^{dy}={}^{x}\zeta^{dy}\,dx+{}^{y}\zeta^{dy}\,dy.
\]
Requiring that equations (\ref{eq:Mar24.2}) are invariants of the
generator $X$:
\[
XS_{i}=0,
\]
one finds that
\[
\begin{array}{c}
\zeta^{dy}=\Delta^{-1}((\zeta^{u}\rho v(p+q_{12}+\rho v^{2})+\zeta^{v}\rho(pu+q_{12}u-2q_{23}v-\rho uv^{2})\\
+\zeta^{\rho}v(pu+q_{12}u-q_{23}v)-\zeta^{p}(q_{23}+\rho uv))dx\\
+(\zeta^{u}\rho(-2pu-2q_{12}u+q_{23}v-\rho uv^{2})+\zeta^{v}\rho u(q_{23}+\rho uv)\\
+\zeta^{\rho}u(-pu-q_{12}u+q_{23}v)-\zeta^{p}(p+q_{12}+\rho v^{2}))dy)
\end{array}
\]
\[
\begin{array}{c}
\zeta^{dx}=\Delta^{-1}((\zeta^{u}\rho v(q_{13}+\rho uv)+\zeta^{v}\rho(-2pv+q_{13}u-2q_{22}v-\rho u^{2}v)\\
+\zeta^{\rho}v(-pv+q_{13}u-q_{22}v)-\zeta^{p}(p+q_{22}+\rho u^{2}))dx\\
+(\zeta^{u}\rho(pv-2q_{13}u+q_{22}v-\rho u^{2}v)+\zeta^{v}\rho u(p+q_{22}+\rho u^{2})\\
+\zeta^{\rho}u(pv-q_{13}u+q_{22}v)-\zeta^{p}(q_{13}+\rho uv))dy),
\end{array}
\]
where
\[
\Delta=p^{2}+(q_{12}+q_{22})p+p\rho(u^{2}+v^{2})+q_{12}q_{22}-q_{13}q_{23}+\rho(q_{12}u^{2}+q_{22}v^{2})-(q_{13}+q_{23})\rho uv.
\]

The next step consists of finding the coefficients $\zeta^{\rho}$,
$\zeta^{u}$, $\zeta^{v}$, $\zeta^{p}$ and $\zeta^{S}$, satisfying
the determining equations
\begin{equation}
(XF_{i})_{|(\ref{eq:2Dgas})}=0,\,\,\,(i=1,2,3,4).\label{eq:determining_2-1}
\end{equation}
Here $X$ is the prolongation of the generator (\ref{eq:apr11.5-1-1})
with the prolongation formulas
\begin{equation}
{\displaystyle \zeta^{f_{x}}=D_{x}\zeta^{f}-f_{x}\frac{\partial\zeta^{dx}}{\partial(dx)}-f_{y}\frac{\partial\zeta^{dy}}{\partial(dx)},\,\,\,\zeta^{f_{y}}=D_{y}\zeta^{f}-f_{x}\frac{\partial\zeta^{dx}}{\partial(dy)}-f_{y}\frac{\partial\zeta^{dy}}{\partial(dy)},}\label{eq:apr11.6-1}
\end{equation}
where $f=\rho,\,\,u,\,\,\,\,v,\,\,\,p,\,\,\,S$, $D_{x}$ and $D_{y}$
are operators of the total derivatives with respect to $x$ and $y$
\[
D_{x}=\partial_{x}+\partial_{\rho}\rho_{x}+\partial_{u}u_{x}+\partial_{v}v_{x}+\partial_{p}p_{x}+\partial_{S}S_{x},
\]
\[
D_{y}=\partial_{y}+\partial_{\rho}\rho_{y}+\partial_{u}u_{y}+\partial_{v}v_{y}+\partial_{p}p_{y}+\partial_{S}S_{y},
\]
the derivatives ${\displaystyle \frac{\partial\zeta^{dx}}{\partial(dx)}}$,
${\displaystyle \frac{\partial\zeta^{dx}}{\partial(dy)}}$, ${\displaystyle \frac{\partial\zeta^{dy}}{\partial(dx)}}$
and ${\displaystyle \frac{\partial\zeta^{dy}}{\partial(dy)}}$ mean
the coefficients of the 1-forms $\zeta^{dx}$ and $\zeta^{dt}$. The
prolongation formulas (\ref{eq:apr11.6-1}) are derived by using the
invariance of the differentials $df$ during the transformations.
Using this invariance, one finds formulas for transformation of derivatives,
and differentiating them with respect to the group-parameter and setting
it to zero, one obtains the prolongation formulas (\ref{eq:apr11.6-1}).

Calculations show that for solving the determining equations, it is
necessary to consider three cases (a) $\zeta^{p}=0$, (b) $\zeta^{p}q_{13}\neq0$,
and (c) $\zeta^{p}\neq0$ and $q_{13}=0$.

The case $\zeta^{p}=0$ gives 
that $X=X_{h}^{e}+X_{F}^{e}$, which means that there are no reciprocal
transformations for this case.

\subsection{Case $\zeta^{p}q_{13}\protect\neq0$}


In this case one obtains that $q_{23}=-q_{13}$, $q_{22}=q_{12}$,
and the general solution of the determining equations (\ref{eq:determining_2-1})
gives that
\[
X=k(X_{3}+2q_{12}X_{4}+q_{13}X_{1}+(q_{12}^{2}+q_{13}^{2})X_{5})+X_{h}^{e}+X_{F}^{e},
\]
where $k$ is an arbitrary constant, $h(S)$ and $F(S)$ are arbitrary
functions. For finding reciprocal transformations the generators $X_{h}^{e}$
and $X_{F}^{e}$ can be excluded. One first notices that $\zeta^{dx}=-k(q_{11}^{-1}S_{1}+2q_{13}dy)$
and $\zeta^{dy}=-k(q_{21}^{-1}S_{2}-2q_{13}dx)$. As $S_{1}$ and
$S_{2}$ are invariants, then $dx^{\prime}=-\epsilon k(q_{11}^{-1}S_{1}+2q_{13}dy)+dx$
and $dy^{\prime}=-\epsilon k(q_{21}^{-1}S_{2}-2q_{13}\,dx)+dy$ or
\begin{equation}
\begin{array}{c}
dx^{\prime}=-\epsilon k\left((p+q_{12}+\rho v^{2})dx-(\rho uv-q_{13})\,dy\right)+dx,\\
dy^{\prime}=\epsilon k\left((\rho uv+q_{13})\,dx-(p+q_{12}+\rho u^{2})dy\right)+dy.
\end{array}\label{eq:jul31.1}
\end{equation}
It is more convenient to obtain transformations of the dependent variables
in cylindrical coordinates for the velocity, defined by the change
\[
u=R\sin(\theta),\,\,\,v=R\cos(\theta).
\]
The solution of the Lie equations becomes
\begin{equation}
\begin{array}{c}
{\displaystyle R^{\prime}=\frac{q_{13}R\sqrt{1+\lambda^{2}}}{q_{13}-\lambda(p+q_{12})},\,\,\,\theta^{\prime}=\theta-\epsilon q_{13},\,\,\,\rho^{\prime}=\rho\frac{\lambda(p+q_{12})-q_{13}}{\lambda(p+q_{12}+\rho R^{2})-q_{13}},}\\[2ex]
{\displaystyle p^{\prime}=\frac{q_{13}p+\lambda(pq_{12}+q_{12}^{2}+q_{13}^{2})}{q_{13}-\lambda(p+q_{12})},\,\,\,S^{\prime}=S,}
\end{array}\label{eq:aug01.4}
\end{equation}
where $\lambda=\tan(\epsilon q_{13})$.

\subsection{Case $\zeta^{p}\protect\neq0$ and $q_{13}=0$}


One has that $q_{23}=0$, $q_{22}=q_{12}$, and the general solution
of the determining equations (\ref{eq:determining_2-1}) gives that
\[
X=k_{2}(X_{3}+2q_{12}X_{4}+q_{12}^{2}X_{5})+k_{1}(2X_{4}+2q_{12}X_{5}-X_{2})+X_{h}^{e}+X_{F}^{e},
\]
where $k_{1}$ and $k_{2}$ are arbitrary constants, $h(S)$ and $F(S)$
are arbitrary functions. One obtains that $\zeta^{dx}=-k_{2}q_{11}^{-1}S_{1}-2k_{1}\,dx$
and $\zeta^{dy}=-k_{2}q_{21}^{-1}S_{2}-2k_{1}\,dy$. As $S_{1}$ and
$S_{2}$ are invariants, then
\begin{equation}
\begin{array}{c}
dx^{\prime}=\epsilon\left(k_{2}\left(-(p+q_{12}+\rho v^{2})dx+\rho uvdy\right)-2k_{1}dx\right)+dx,\\[2ex]
dy^{\prime}=\epsilon\left(k_{2}\left(\rho uvdx-(p+q_{12}+\rho u^{2})dy\right)-2k_{1}dy\right)+dy.
\end{array}\label{eq:aug01.5}
\end{equation}

The formulas for changing the dependent variables depend on $k_{1}$.

For $k_{1}\neq0$, one finds
\begin{equation}
\begin{array}{c}
{\displaystyle u^{\prime}=-\frac{2k_{1}u\lambda}{k_{2}(\lambda^{2}-1)(p+q_{12})-2k_{1}},\,\,\,v^{\prime}=-\frac{2k_{1}v\lambda}{k_{2}(\lambda^{2}-1)(p+q_{12})-2k_{1}},}\\[2ex]
{\displaystyle \rho^{\prime}=\rho\frac{k_{2}(\lambda^{2}-1)(p+q_{12})-2k_{1}}{k_{2}(\lambda^{2}-1)(p+q_{12}+\rho(u^{2}+v^{2}))-2k_{1}},}\\[2ex]
{\displaystyle p^{\prime}=\frac{k_{2}q_{12}(p+q_{12})(1-\lambda^{2})+2k_{1}(q_{12}(1-\lambda^{2})-\lambda^{2}p)}{k_{2}(p+q_{12})(\lambda^{2}-1)-2k_{1}},}
\end{array}\label{eq:aug01.1}
\end{equation}
where $\lambda=e^{k_{1}\epsilon}$.

For $k_{1}=0$, one has 
\begin{equation}
\begin{array}{c}
{\displaystyle u^{\prime}=\frac{u}{1-a(p+q_{12})},\,\,\,v^{\prime}=\frac{v}{1-a(p+q_{12})},}\\[2ex]
{\displaystyle \rho^{\prime}=\rho\frac{1-a(p+q_{12})}{1-a(p+q_{12}+\rho(u^{2}+v^{2}))},\,\,\,p^{\prime}=\frac{p-aq_{12}(p+q_{12})}{1-a(p+q_{12})},}
\end{array}\label{eq:aug01.2}
\end{equation}
where $a=k_{2}\epsilon$.

\section{Application of the second method}

In the previous sections the differential forms $S_{1}$ and $S_{2}$
are used for constructing reciprocal transformations, while this section
shows that they are not needed. In the second method there is no separation
in the two steps like in the first method, as the second method does
not use the conservation laws. The prolongation formulas of the generator
of the group of reciprocal transformations are constructed by using
the general form of the coefficients related with the differentials
of the generator of the group of reciprocal transformations. This
method is a little bit more complicated, but it can also be applied
to systems without knowing their conservation laws.

\subsection{Group of reciprocal transformations of (\ref{eq:2Dgas}) }

Consider the generator

\begin{equation}
X=\zeta^{\rho}\partial_{\rho}+\zeta^{v}\partial_{v}+\zeta^{p}\partial_{p}+\zeta^{S}\partial_{S}+\zeta^{dt}\partial_{dt}+\zeta^{dx}\partial_{dx},\label{eq:apr11.5-1}
\end{equation}
where the coefficients $\zeta^{\rho}$, $\zeta^{u}$, $\zeta^{v}$,
$\zeta^{p}$ and $\zeta^{S}$ depend on the variables $(x,y,\rho,u,v,p,S)$,
$\zeta^{dx}$ and $\zeta^{dy}$ are linear 1-forms with respect to
$dx$ and $dy$ with the coefficients also depending on the variables
$(x,y,\rho,u,v,p,S)$:
\[
\zeta^{dx}={}^{x}\zeta^{dx}\,dx+{}^{y}\zeta^{dx}\,dy,\,\,\,\zeta^{dy}={}^{x}\zeta^{dy}\,dx+{}^{y}\zeta^{dy}\,dy.
\]
Requiring that equations (\ref{eq:2Dgas}) compose an invariant manifold
of the generator $X$, one derives the determining equations

\begin{equation}
\begin{array}{c}
\left(\rho\zeta_{x}^{u}+\rho\zeta_{y}^{v}+\rho_{x}\zeta^{u}+\rho_{y}\zeta^{v}+u\zeta^{\rho_{x}}+u_{x}\zeta^{\rho}+v\zeta^{\rho_{y}}+v_{y}\zeta^{\rho}\right)_{|(\ref{eq:2Dgas})}=0,\\[2ex]
\left(\rho u\zeta_{x}^{u}+\rho u_{x}\zeta^{u}+\rho u_{y}\zeta^{v}+\rho v\zeta_{y}^{u}+uu_{x}\zeta^{\rho}+u_{y}v\zeta^{\rho}+\zeta_{x}^{p}\right)_{|(\ref{eq:2Dgas})}=0,\\[2ex]
\left(\rho u\zeta_{x}^{v}+\rho v\zeta_{y}^{v}+\rho v_{x}\zeta^{u}+\rho v_{y}\zeta^{v}+uv_{x}\zeta^{\rho}+vv_{y}\zeta^{\rho}+\zeta_{y}^{p}\right)_{|(\ref{eq:2Dgas})}=0,\\[2ex]
\left(S_{x}\zeta^{u}+S_{y}\zeta^{v}+u\zeta_{x}^{S}+v\zeta_{y}^{S}\right)_{|(\ref{eq:2Dgas})}=0,
\end{array}\label{eq:jul12.1}
\end{equation}
where the sign $_{|(\ref{eq:2Dgas})}$ has the usual meaning of considering
the relations in parenthesis on the manifold defined by equations
(\ref{eq:2Dgas}), where for $\zeta^{f_{x}}$ and $\zeta^{f_{y}}$,
$(f=\rho,\,u,\,v,p\,,\,S)$ one has to apply the prolongation formulas
(\ref{eq:apr11.6-1}).

Beside equations (\ref{eq:jul12.1}) one also has to satisfy the equations
$D_{x}\left(^{y}\zeta^{dx}\right)-D_{y}\left(^{x}\zeta^{dx}\right)=0$
and $D_{x}\left(^{y}\zeta^{dy}\right)-D_{y}\left(^{x}\zeta^{dy}\right)=0$,
which are

\begin{equation}
\begin{array}{c}
-^{x}\zeta_{S}^{dx}S_{y}-{}^{x}\zeta_{p}^{dx}p_{y}-{}^{x}\zeta_{\rho}^{dx}\rho_{y}-{}^{x}\zeta_{u}^{dx}u_{y}-{}^{x}\zeta_{v}^{dx}v_{y}-{}^{x}\zeta_{y}^{dx}\\
+^{y}\zeta_{S}^{dx}S_{x}+{}^{y}\zeta_{p}^{dx}p_{x}+{}^{y}\zeta_{\rho}^{dx}\rho_{x}+{}^{y}\zeta_{u}^{dx}u_{x}+{}^{y}\zeta_{v}^{dx}v_{x}+{}^{y}\zeta_{x}^{dx}=0,\\[2ex]
-^{x}\zeta_{S}^{dy}S_{y}-{}^{x}\zeta_{p}^{dy}p_{y}-{}^{x}\zeta_{\rho}^{dy}\rho_{y}-{}^{x}\zeta_{u}^{dy}u_{y}-{}^{x}\zeta_{v}^{dy}v_{y}-{}^{x}\zeta_{y}^{dy}\\
+^{y}\zeta_{S}^{dy}S_{x}+{}^{y}\zeta_{p}^{dy}p_{x}+{}^{y}\zeta_{\rho}^{dy}\rho_{x}+{}^{y}\zeta_{u}^{dy}u_{x}+{}^{y}\zeta_{v}^{dy}v_{x}+{}^{y}\zeta_{x}^{dy}=0.
\end{array}\label{eq:jul12.2}
\end{equation}

The method of solving the determining equations (\ref{eq:jul12.1}),
(\ref{eq:jul12.2}) is similar as in the classical group analysis
method \citep{bk:Ovsiannikov1978}. Calculations show that, solving
the determining equations (\ref{eq:jul12.1}), (\ref{eq:jul12.2}),
one obtains that
\[
X=\sum_{i=2}^{5}k_{i}X_{i}+X_{h}^{e}+X_{F}^{e},
\]
where and $h(S)$ and $F(S)$ are arbitrary functions, $k_{i}$, ($i=1,2,...,5$)
are arbitrary constants, and the generators $X_{i}$, ($i=1,2,...,5$)
are defined by formulas (\ref{eq:aug01.10}):
\begin{equation}
\begin{array}{c}
X_{1}=-v\partial_{u}+u\partial_{v}-dy\partial_{dx}+dx\partial_{dy},\,\,\,X_{2}=dx\partial_{dx}+dy\partial_{dy},\\[2ex]
X_{3}=\rho^{2}q^{2}\partial_{\rho}+pu\partial_{u}+pv\partial_{v}+p^{2}\partial_{p}+\left(-(p+\rho v^{2})dx+\rho uvdy\right)\partial_{dx}\\
+\left(\rho uvdx-(p+\rho u^{2})dy\right)\partial_{dy},\\[2ex]
X_{4}=\frac{1}{2}\left(2p\partial_{p}+u\partial_{u}+v\partial_{v}-dx\partial_{dx}-dy\partial_{dy}\right),\,\,\,X_{5}=\partial_{p}.
\end{array}\label{eq:aug01.11}
\end{equation}

\subsection{Algebraic properties of the Lie algebra $L_{rt}=\{X_{1},X_{2},X_{3},X_{4},X_{5},X_{h},X_{F}\}$ }

The commutator table of $L_{rt}$ is
\[
\begin{array}{c||cc|ccc|cc}
 & X_{1} & X_{2} & X_{3} & X_{4} & X_{5} & X_{h} & X_{F}\\
\hline\hline X_{1} & 0 & 0 & 0 & 0 & 0 & 0 & 0\\
X_{2} & 0 & 0 & 0 & 0 & 0 & 0 & 0\\
\hline X_{3} & 0 & 0 & 0 & -X_{3} & -X_{4} & 0 & 0\\
X_{4} & 0 & 0 & X_{3} & 0 & -X_{5} & 0 & 0\\
X_{5} & 0 & 0 & X_{4} & X_{5} & 0 & 0 & 0\\
\hline X_{h} & 0 & 0 & 0 & 0 & 0 & 0 & -X_{h_{1}}\\
X_{F} & 0 & 0 & 0 & 0 & 0 & X_{h_{1}} & 0
\end{array}
\]
where $h_{1}=h^{\prime}F$. From the commutator table, one concludes
that the generators $X_{1}$ and $X_{2}$ compose the center, the
generators $X_{3}$, $X_{4}$ and $X_{5}$ compose an ideal, the derivatives
of $L_{rt}$ are
\[
L_{rt}^{\prime}=[L_{rt},L_{rt}]=\{X_{3},X_{4},X_{5},X_{h}\},\,\,\,L_{rt}^{\prime\prime}=[L_{rt}^{\prime},L_{rt}^{\prime}]=\{X_{3},X_{4},X_{5}\}.
\]

\section{Complete set of reciprocal transformations}

For finding discrete reciprocal transformations, it is adapted the
method proposed in \citep{art:Hydon1998} for finding discrete symmetries,
and extended in \citep{bk:BihloBihloPopovych2015} for equivalence
transformations. The main idea of the method consists of using invariant
sets of a group of reciprocal transformations.

Let be given a system of differential equations $(S)$, and $L$ be
its (maximal) Lie algebra of reciprocal transformations, where $Aut(L)$
is the automorphism group. Any reciprocal transformation $T$ provides
an automorphism $T_{*}\in Aut(L)$ of the Lie algebra $L$. The transformation
$T_{*}$ is defined by the change of a generator under the transformation
$T$. Notice that the change of the generator of a group of reciprocal
transformations under a reciprocal transformation follows the same
formulas as the change of the generator of a group of point transformations
under a point transformation \citep{bk:Ovsiannikov1978}. The property
that $T_{*}\in Aut(L)$ follows from commutativeness of the operation
of commutation of generators and the operation of the change of the
variables.

For further study we use the following notations\footnote{See details in \citep{bk:BihloBihloPopovych2015} and references therein. }.
A megaideal $\tau$ of a Lie algebra $L$ is a vector subspace of
$L$, which is invariant under any mapping from the automorphism group
$Aut(L)$. That is, for any megaideal $\tau$ of $L$, and any transformation
$A$ from $Aut(L)$ one has that $A\tau=\tau$.

Some megaideals of $L$ can be computed without knowing $Aut(L)$.
In particular, the center and the derivative of a Lie algebra are
its megaideals \citep{bk:BihloBihloPopovych2015}. Hence, $L^{\prime}=[L,L]$
is a megaideal. We also use the property that a megaideal $\tau_{2}$
of a megaideal $\tau_{1}$ of $L$ is also a megaideal of $L$. In
particular, $L^{\prime\prime}=(L^{\prime})^{\prime}$ is also a megaideal
\citep{bk:BihloBihloPopovych2015}.

Thus, one obtains the megaideals of $L_{rt}$ :
\[
L_{rt}^{\prime}=\{X_{3},X_{4},X_{5},X_{h}\},\,\,\,L_{rt}^{\prime\prime}=\{X_{3},X_{4},X_{5}\},\,\,\,C_{1}=\{X_{1}\},\,\,\,C_{2}=\{X_{2}\}.
\]
Notice also that the nonzero structure constants for $L_{rt}^{\prime\prime}$
are
\[
c_{34}^{3}=-1,\,\,\,c_{35}^{4}=-1,\,\,\,c_{45}^{5}=-1.
\]

Consider an invertible transformation $T$ defined by the relations
\begin{equation}
\begin{array}{c}
\rho^{\prime}=R(\rho,u,v,p,S),\,\,\,u^{\prime}=U(\rho,u,v,p,S),\,\,\,v^{\prime}=V(\rho,u,v,p,S),\\[2ex]
p^{\prime}=P(\rho,u,v,p,S),\,\,\,S^{\prime}=H(\rho,u,v,p,S),\\[2ex]
dx^{\prime}={}^{x}f^{dx}(\rho,u,v,p,S)\,dx+{}^{y}f^{dx}(\rho,u,v,p,S)\,dy,\\
dy^{\prime}={}^{x}f^{dy}(\rho,u,v,p,S)\,dx+{}^{y}f^{dy}(\rho,u,v,p,S)\,dy.
\end{array}\label{eq:aug03.1}
\end{equation}
The transformation $T$ is a reciprocal transformation if
\begin{equation}
\begin{array}{c}
-RUdy^{\prime}+RVdx^{\prime}=0,\,\,\,(P+RV^{2})dx^{\prime}-RUVdy^{\prime}=0,\\
RUVdx^{\prime}-(RU{}^{2}+P)dy^{\prime}=0,\,\,\,RHUdy^{\prime}-RVHdx^{\prime}=0.
\end{array}\label{eq:recip}
\end{equation}

The transformation $T_{*}$ acts on a generator $X\in L_{rt}$
\begin{equation}
X=\zeta^{\rho}\partial_{\rho}+\zeta^{v}\partial_{v}+\zeta^{p}\partial_{p}+\zeta^{S}\partial_{S}+\zeta^{dt}\partial_{dt}+\zeta^{dx}\partial_{dx},\label{eq:jul19.1}
\end{equation}
as follows
\begin{equation}
T_{*}X=\zeta^{\rho^{\prime}}\partial_{\rho^{\prime}}+\zeta^{v^{\prime}}\partial_{v^{\prime}}+\zeta^{p^{\prime}}\partial_{p^{\prime}}+\zeta^{S^{\prime}}\partial_{S^{\prime}}+\zeta^{dt^{\prime}}\partial_{dt^{\prime}}+\zeta^{dx^{\prime}}\partial_{dx^{\prime}},\label{eq:jul19.2}
\end{equation}
where
\[
\zeta^{\rho^{\prime}}=X\rho^{\prime},\,\,\,\zeta^{v^{\prime}}=Xv^{\prime},\,\,\,\zeta^{p^{\prime}}=Xp^{\prime},\,\,\,\zeta^{S^{\prime}}=XS^{\prime},\,\,\,\zeta^{dt^{\prime}}=X(dt^{\prime}),\,\,\,\zeta^{dx^{\prime}}=X(dx^{\prime}).
\]

One of the sets of equations for the invertible transformation $T$
to be a reciprocal transformation is defined by the conditions that
$dx^{\prime}$ and $dy^{\prime}$ are differentials:
\begin{equation}
D_{y}(^{x}f^{dx})=D_{x}({}^{y}f^{dx}),\,\,\,D_{y}({}^{x}f^{dy})=D_{x}({}^{y}f^{dy}).\label{eq:jul20.1}
\end{equation}
Analysis of these equations is similar to the analysis of determining
equations. As these equations have to be satisfied for any solution
of equations (\ref{eq:2Dgas}), then substituting the main derivatives
of equations (\ref{eq:2Dgas}) into (\ref{eq:jul20.1}), and splitting
them with respect to parametrical derivatives, one derives the set
of equations, which we also call determining equations.

\subsection{Use of the megaideal $L_{rt}^{\prime\prime}$ }

Let $X$ belong to the megaideal $L_{rt}^{\prime\prime}=\{X_{3},X_{4},X_{5}\}.$
As $T_{*}\in Aut(L_{rt}^{\prime\prime})$, then $T_{*}X\in L_{rt}^{\prime\prime}$,
which means that there exist constants such that
\[
T_{*}X=a_{3}X_{3}^{\prime}+a_{4}X_{4}^{\prime}+a_{5}X_{5}^{\prime},
\]
where prime in $X_{i}^{\prime}$ means that it is the generator $X_{i}$
with primed variables:
\[
\begin{array}{c}
X_{3}^{\prime}=\rho^{\prime\,2}q^{\prime\,2}\partial_{\rho^{\prime}}+p^{\prime}u^{\prime}\partial_{u^{\prime}}+p^{\prime}v^{\prime}\partial_{v^{\prime}}+p^{\prime\,2}\partial_{p^{\prime}}\\[2ex]
+\left(-(p^{\prime}+\rho^{\prime}v^{\prime\,2})dx^{\prime}+\rho^{\prime}u^{\prime}v^{\prime}dy^{\prime}\right)\partial_{dx^{\prime}}+\left(\rho^{\prime}u^{\prime}v^{\prime}dx^{\prime}-(p^{\prime}+\rho^{\prime}u^{\prime\,2})dy^{\prime}\right)\partial_{dy^{\prime}},
\end{array}
\]
\[
X_{4}^{\prime}=\frac{1}{2}\left(2p^{\prime}\partial_{p^{\prime}}+u^{\prime}\partial_{u^{\prime}}+v^{\prime}\partial_{v^{\prime}}-dx^{\prime}\partial_{dx^{\prime}}-dy^{\prime}\partial_{dy^{\prime}}\right),\,\,\,X_{5}^{\prime}=\partial_{p^{\prime}}.
\]
In particular,
\begin{equation}
T_{*}X_{i}=a_{3i}X_{3}^{\prime}+a_{4i}X_{4}^{\prime}+a_{5i}X_{5}^{\prime},\,\,\,(i=3,4,5),\label{eq:jul19.5}
\end{equation}
where the matrix
\[
A=\left(\begin{array}{ccc}
a_{33} & a_{34} & a_{35}\\
a_{43} & a_{44} & a_{45}\\
a_{53} & a_{54} & a_{55}
\end{array}\right)
\]
is nonsingular.

The constants $a_{ij},\,\,\,(i,j=3,4,5)$ have to satisfy the following
relations. As $T_{*}\in Aut(L_{rt}^{\prime\prime})$, then
\[
[T_{*}X_{i},T_{*}X_{j}]=T_{*}[X_{i},X_{j}]=\sum_{k=3}^{5}c_{ij}^{k}T_{*}X_{k},\,\,\,(i,j=3,4,5),
\]
which lead to the conditions
\[
\sum_{k=3}^{5}\,\sum_{s=3}^{5}\,\sum_{n=3}^{5}a_{ik}a_{js}c_{ks}^{n}X_{n}^{\prime}=\sum_{k=3}^{5}\,\sum_{n=3}^{5}c_{ij}^{k}a_{nk}X_{n}^{\prime},\,\,\,(i,j=3,4,5)
\]
or
\begin{equation}
\sum_{k=3}^{5}\,\sum_{s=3}^{5}\,a_{ik}a_{js}c_{ks}^{n}=\sum_{k=3}^{5}\,c_{ij}^{k}a_{nk},\,\,\,(i,j,n=3,4,5)\label{eq:jul19.6}
\end{equation}

Calculations show that equations (\ref{eq:jul19.5}) define the derivatives
with respect $\rho$, $u$ and $p$ of all unknown functions (\ref{eq:aug03.1}).
The solution of equations (\ref{eq:jul19.5}) and the representation
of equations (\ref{eq:jul19.6}) are given in the Appendix. Analysis
of equations (\ref{eq:jul19.6}) leads to the study of two cases (a)
$a_{35}=0$, and (b) $a_{35}\neq0$.

All calculations were performed using symbolic manipulation system
Reduce \citep{bk:Hearn}.

\subsection{Case $a_{35}=0$}

In this case, by virtue of $\det A\neq0$, one derives from equations
(\ref{eq:jul19.6}) that $a_{33}\neq0$ and
\begin{equation}
a_{34}=0,\,\,\,a_{43}=a_{54}a_{33},\,\,\,a_{44}=1,\,\,\,a_{45}=0,\,\,\,a_{53}=\frac{1}{2}a_{54}^{2}a_{33},\,\,\,a_{55}=a_{33}^{-1}.\label{eq:case_a35_eq0}
\end{equation}

\subsubsection{Using the megaideal $\{X_{1}\}$ }

The generator $X_{1}$ is a center of the Lie algebra $L$. As any
center is a megaideal, then there exists constant $a_{11}$ such that
\begin{equation}
T_{*}X_{1}=a_{11}X_{1}^{\prime}.\label{eq:jul21.1}
\end{equation}
Calculations give that the latter equation defines derivatives with
respect $v$ of all unknown functions:

\begin{equation}
R_{v}=0,\,\,\,U_{v}=\frac{Uv-Va_{11}u}{u^{2}+v^{2}},\,\,\,V_{v}=\frac{Ua_{11}u+Vv}{u^{2}+v^{2}},\,\,\,P_{v}=\frac{2Pa_{33}v+2v(a_{54}a_{33}-p)}{a_{33}(u^{2}+v^{2})},\,\,\,H_{v}=0,\label{eq:jul24.3}
\end{equation}

\[
^{x}f_{v}^{dx}=-u\frac{^{y}f^{dx}+{}^{x}f^{dy}a_{11}}{u^{2}+v^{2}},\,\,\,{}^{y}f_{v}^{dx}=u\frac{^{x}f^{dx}-{}^{y}f^{dy}a_{11}}{u^{2}+v^{2}},
\]

\[
^{x}f_{v}^{dy}=u\frac{^{x}f^{dx}a_{11}-{}^{y}f^{dy}}{u^{2}+v^{2}},\,\,\,{}^{y}f_{v}^{dy}=u\frac{^{y}f^{dx}a_{11}+{}^{x}f^{dy}}{u^{2}+v^{2}}.
\]
The remaining equations are equations (\ref{eq:jul20.1}).

\subsubsection{Analysis of equations (\ref{eq:jul20.1})}

Splitting these equations with respect to parametric derivatives of
equations (\ref{eq:2Dgas}), one obtains the determining equations

\begin{equation}
a_{11}^{2}=1,\,\,\,{}^{y}f^{dx}=-{}^{x}f^{dy}a_{11},\,\,\,{}^{x}f^{dx}={}^{y}f^{dy}a_{11},\label{eq:jul24.10}
\end{equation}

\begin{equation}
\begin{array}{c}
{}^{x}f^{dy}a_{11}(p+\rho u^{2}-a_{33}(P+RV{}^{2}+a_{54}))+{}^{y}f^{dy}(a_{11}\rho uv-RUVa_{33})=0,\\[2ex]
^{x}f^{dy}(a_{11}\rho uv+RUVa_{33})+{}^{y}f^{dy}a_{11}(p+\rho v^{2}-a_{33}(P+RV{}^{2}+a_{54}))=0,
\\[2ex]
-^{x}f^{dy}a_{11}(RUVa_{33}+a_{11}\rho uv)+{}^{y}f^{dy}(p+\rho u^{2}-a_{33}(P+RU{}^{2}+a_{54}))=0,
\\[2ex]
-^{x}f^{dy}a_{11}(p+\rho v^{2}-a_{33}(P+RU{}^{2}+a_{54}))+{}^{y}f^{dy}(a_{11}\rho uv-RUVa_{33})=0,
\end{array}\label{eq:fiu(50,2)}
\end{equation}

Considering equations (\ref{eq:fiu(50,2)}) as a system of linear
algebraic equations with respect to $^{x}f^{dy}$ and $^{y}f^{dy}$,
and because $(^{x}f^{dy})^{2}+(^{y}f^{dy})^{2}\neq0$, one derives
that

\begin{equation}
P=\frac{1}{2}\left(\frac{2p+\rho(u^{2}+v^{2})}{a_{33}}-R(U{}^{2}+V{}^{2}+2a_{54})\right),\,\,\,R=\mu\rho\frac{u^{2}+v^{2}}{a_{33}(U{}^{2}+V{}^{2})},\label{eq:jul24.1}
\end{equation}
where $\mu^{2}=1$, and equations (\ref{eq:fiu(50,2)}) are reduced
to the equations
\begin{equation}
\begin{array}{c}
\mu\left(^{x}f^{dy}(V{}^{2}-U{}^{2})+2\,{}^{y}f^{dy}UVa_{11}\right)(u^{2}+v^{2})\\
+\left(-2\,^{y}f^{dy}uv+{}^{x}f^{dy}(v^{2}-u^{2})\right)(V{}^{2}+U{}^{2})=0,\\[2ex]
\left(-2^{x}f^{dy}uv+{}^{y}f^{dy}(u^{2}-v^{2})\right)(U{}^{2}+V{}^{2})\\
+\mu\left(^{y}f^{dy}(V{}^{2}-U{}^{2})-2{}^{x}f^{dy}UVa_{11}\right)(u^{2}+v^{2})=0.
\end{array}\label{eq:jul24.2}
\end{equation}

Integrating equations (\ref{eq:jul19.5})\footnote{Presented in Appendix.}
and (\ref{eq:jul24.3}) for $U$ and $V$, one derives that
\begin{equation}
U=\rho^{(1-\mu)/2}\left(\varphi_{1}u+\varphi_{2}v\right),\,\,\,V=a_{11}\rho^{(1-\mu)/2}(\varphi_{1}v-\varphi_{2}u),\label{eq:jul24.31}
\end{equation}
where $\varphi_{1}(S)$ and $\varphi_{2}(S)$ are arbitrary functions
such that $\varphi_{1}^{2}+\varphi_{2}^{2}\neq0$. Equations (\ref{eq:jul24.1})
become
\begin{equation}
P=\frac{p}{a_{33}}-a_{54}+\frac{(1-\mu)}{2a_{33}}\rho(u^{2}+v^{2}),\,\,\,R=\frac{\mu\rho^{\mu}}{a_{33}(\varphi_{1}^{2}+\varphi_{2}^{2})}.\label{eq:jul24.32}
\end{equation}
Equations (\ref{eq:jul24.10}) give
\begin{equation}
dx^{\prime}=a_{11}({}^{y}f^{dy}dx-{}^{x}f^{dy}dy),\,\,\,dy^{\prime}={}^{x}f^{dy}dx+{}^{y}f^{dy}dy.\label{eq:jul24.33}
\end{equation}

Further analysis of finding the coefficients $^{x}f^{dy}$ and $^{y}f^{dy}$depends
on $\mu$.

If $\mu=1$, then equations (\ref{eq:jul24.2}) provide that

\[
^{x}f^{dy}\varphi_{1}+{}^{y}f^{dy}\varphi_{2}=0.
\]
In symmetric form one can represent a solution of the latter equation
as

\[
^{y}f^{dy}=\varphi_{1}\psi^{-1},\,\,\,{}^{x}f^{dy}=-\varphi_{2}\psi^{-1},
\]
where it can it is also obtained that $\psi(S)$ is an arbitrary function.
Equations (\ref{eq:jul20.1}) require that
\[
\varphi_{1}=\alpha\psi,\,\,\,\varphi_{2}=\beta\psi,
\]
where $\alpha$ and $\beta$ are constant such that $\alpha^{2}+\beta^{2}\neq0$.

Thus, one obtains that the transformation can be written in the form
\begin{equation}
\begin{array}{c}
{\displaystyle P=\frac{p}{a_{33}}-a_{54},\,\,\,R=\frac{\rho}{a_{33}\psi^{2}(\alpha{}^{2}+\beta^{2})},\,\,\,U=\psi(\alpha u+\beta v),\,\,\,V=a_{11}\psi(\alpha v-\beta u),\,\,\,H=F(S),}\\[2ex]
dx^{\prime}=a_{11}(\alpha dx+\beta dy),\,\,\,dy^{\prime}=-\beta dx+\alpha dy.
\end{array}\label{eq:jul24.11}
\end{equation}
As the coefficients $^{x}f^{dx}$, $^{y}f^{dx}$, $^{x}f^{dy}$ and
$^{y}f^{dy}$ are constant, then the transformation (\ref{eq:jul24.11})
is equivalent to an equivalence transformation: in particular, composition
of (\ref{eq:aug02.1}) with the rotation.

If $\mu=-1$, then (\ref{eq:jul24.31}), (\ref{eq:jul24.32}) become
\begin{equation}
P=\frac{p}{a_{33}}-a_{54}+\frac{1}{a_{33}}\rho(u^{2}+v^{2}),\,\,\,R=-\frac{1}{a_{33}\rho(\varphi_{1}^{2}+\varphi_{2}^{2})},\label{eq:jul24.34}
\end{equation}
\begin{equation}
U=\rho\left(\varphi_{1}u+\varphi_{2}v\right),\,\,\,V=a_{11}\rho(\varphi_{1}v-\varphi_{2}u).\label{eq:jul24.31-1}
\end{equation}
Equations (\ref{eq:jul24.2}) give that
\[
^{x}f^{dy}\varphi_{2}-{}^{y}f^{dy}\varphi_{1}=0,
\]
In symmetric form one can represent a solution of the latter equation
as $^{y}f^{dy}=\varphi_{2}\psi^{-1},\,\,\,{}^{x}f^{dy}=\varphi_{1}\psi^{-1}.$
Equations (\ref{eq:jul20.1}) require that $\varphi_{1}=\alpha\psi,\,\,\,\varphi_{2}=\beta\psi$.
Equations (\ref{eq:recip}) in this case are only satisfied if the
original solution is isentropic and irrotational. Hence, the case
$\mu=-1$ also does not provide a reciprocal transformation.

\subsection{Case $a_{35}\protect\neq0$ }

For this case
\[
a_{33}=\frac{a_{34}^{2}}{2a_{35}},\,\,\,a_{43}=\frac{a_{34}(a_{45}a_{34}-2a_{35})}{2a_{35}^{2}},\,\,\,a_{44}=\frac{a_{45}a_{34}}{a_{35}}-1,
\]
\[
a_{53}=\frac{a_{45}^{2}a_{34}^{2}-4a_{45}a_{35}a_{34}+4a_{35}^{2}}{4a_{35}^{3}},\,\,\,a_{54}=\frac{a_{45}(a_{45}a_{34}-2a_{35})}{2a_{35}^{2}},\,\,\,a_{55}=\frac{a_{45}^{2}}{2a_{35}},
\]
and analysis of the equations defining the reciprocal transformations
is similar to the case $a_{35}=0$, but more cumbersome. Because of
their cumbersomeness, we only describe the main steps of the finding
of the reciprocal transformations.

Using the megaideal $\{X_{1}\}$, one finds the derivatives $R_{v}$,
$U_{v}$, $V_{v}$, $P_{v}$, $^{x}f_{v}^{dx}$, $^{y}f_{v}^{dx}$,
$^{x}f_{v}^{dy}$, and $^{y}f_{v}^{dy}$. After that equations (\ref{eq:jul20.1})
give the relation $u\,{}^{x}f_{S}^{dx}+v\,{}^{y}f_{S}^{dx}=0$, and
an algebraic system of ten homogeneous linear equations with respect
to $^{x}f^{dx}$, $^{y}f^{dx}$, $^{x}f^{dy}$, and $^{y}f^{dy}$.
As $(^{x}f^{dx})({}^{y}f^{dy})-({}^{y}f^{dx})({}^{x}f^{dy})\ne0$,
then the rank $r$ of the matrix with respect to these variables satisfies
the inequality $r\leq3$. From the analysis of the minors of latter
system of linear homogeneous equations one finds $P$ and $R$. Integrating
the overdetermined system of equations for the functions $U$ and
$V$, one funds them. Substituting all the expressions of $R$, $U$,
$V$, and $P$ into a linear system for $^{x}f^{dx}$, $^{y}f^{dx}$,
$^{x}f^{dy}$, and $^{y}f^{dy}$, one finds the reciprocal transformations
\begin{equation}
\begin{array}{c}
{\displaystyle R=\frac{2\rho(p-g)}{\psi^{2}a_{35}(\alpha{}^{2}+\beta^{2})(p+\rho q^{2}-g)},\,\,\,P=-\frac{a_{45}}{a_{35}}-\frac{2}{a_{35}(p-g)},}\\[2ex]
{\displaystyle U=\frac{\psi(\alpha v+\beta u)}{p-g},\,\,\,V=a_{11}\frac{\psi(-\alpha u+\beta v)}{p-g},\,\,\,H=F,}\\[2ex]
{\displaystyle dx^{\prime}=k\left((\alpha\rho uv-\beta(p+\rho v^{2}-g))\,dx+(-\alpha(p+\rho u^{2}-g)+\beta\rho uv)\,dy\right),}\\[2ex]
{\displaystyle dy^{\prime}=ka_{11}\left((\alpha(p+\rho v^{2}-g)+\beta\rho uv)\,dx-(\alpha\rho uv+\beta(p+\rho u^{2}-g))\,dy\right),}
\end{array}\label{eq:aug01.20}
\end{equation}
where $g=a_{34}a_{35}^{-1}$, $a_{11}^{2}=1$, $\alpha$, $\beta$,
and $k$ are constant, $\psi(S)$ and $F(S)$ are arbitrary functions.
Recall that $a_{34}$, and $a_{45}$ are arbitrary constants, and
$a_{35}\neq0$.

Notice that because of the equivalence transformation corresponding
to the involution $E_{2}$ and the rotation, one can assume that $a_{11}=1$
and\footnote{As $\alpha^{2}+\beta^{2}\neq0$, one has $\beta\neq0$. }
$\alpha=0$. By virtue of the equivalence transformation (\ref{eq:aug02.1}),
one can reduce $\psi$ from formulas (\ref{eq:aug01.20}). Introducing
the constants $\beta_{i},\,\,(i=1,2,3,4)$:

\[
\beta=\frac{2}{\beta_{1}\beta_{3}},\,\,\,a_{34}=-\frac{2\beta_{2}}{\beta_{1}^{2}\beta_{3}},\,\,\,a_{35}=\frac{2}{\beta_{1}^{2}\beta_{3}},\,\,\,a_{45}=-\frac{2\beta_{4}}{\beta_{1}^{2}\beta_{3}},\,\,\,k=-\frac{\beta_{3}}{2},
\]
formulas (\ref{eq:aug01.20}) coincide with (\ref{eq:2.4}), (\ref{eq:2.6}).

From the above study one can conclude that for finding all reciprocal
transformations it was sufficient to use the megaideal $L_{tr}^{\prime\prime}=\{X_{3},X_{4},X_{5}\}$
and the center $\{X_{1}\}$. The final results obtained can be formulated
as follows.

\begin{theorem} The complete set of reciprocal transformations of
the two-dimensional stationary gas dynamics equations, considered
up to the equivalence transformations corresponding to (\ref{eq:may06-1})
and the involution $E_{2}$, consists of the transformations (\ref{eq:2.4}),
(\ref{eq:2.6}):
\begin{equation}
\begin{array}{c}
{\displaystyle u^{\prime}=\frac{\beta_{1}u}{p+\beta_{2}},\,\,\ v^{\prime}=\frac{\beta_{1}v}{p+\beta_{2}},}\\
{\displaystyle p^{\prime}=\beta_{4}-\frac{\beta_{1}^{2}\beta_{3}}{p+\beta_{2}},\,\,\ \rho^{\prime}=\frac{\beta_{3}\rho(p+\beta_{2})}{p+\beta_{2}+\rho q^{2}},\,\,\ S^{\prime}=F(S),}
\end{array}\label{eq:2.6-2}
\end{equation}
\begin{equation}
\begin{array}{c}
dx^{\prime}=\beta_{1}^{-1}[(p+\beta_{2}+\rho v^{2})dx-\rho uvdy],\\
dy^{\prime}=\beta_{1}^{-1}[-\rho uvdx+(p+\beta_{2}+\rho u^{2})dy].
\end{array}\label{eq:2.4-2}
\end{equation}
\end{theorem}

\begin{remark}The transformations (\ref{eq:2.4-2}), (\ref{eq:2.6-2})
can be further simplified by the equivalence transformations corresponding
to (\ref{eq:may06-1}). In particular, using the transformation corresponding
to $X_{6}^{e}$, one can assume that $\beta_{2}=0$. Because of the
transformation corresponding to $X_{5}^{e}$, one can assume that
$\beta_{1}=1$. The reciprocal transformations (\ref{eq:2.4-2}),
(\ref{eq:2.6-2}) become
\[
\begin{array}{c}
{\displaystyle u^{\prime}=\frac{u}{p},\,\,\ v^{\prime}=\frac{v}{p},\,\,\,p^{\prime}=\beta_{4}-\frac{\tilde{\beta}_{3}}{p},\,\,\ \rho^{\prime}=\frac{\tilde{\beta}_{3}\rho p}{p+\rho q^{2}},\,\,\ S^{\prime}=F(S),}\\[2ex]
dx^{\prime}=(p+\rho v^{2})dx-\rho uvdy,\,\,\,dy^{\prime}=-\rho uvdx+(p+\rho u^{2})dy.
\end{array}
\]
\end{remark}

\section{Conclusions}

Two methods for constructing a group of reciprocal transformations
are presented in the paper. These methods are demonstrated by the
two-dimensional stationary gas dynamics equations. Both methods use
the infinitesimal approach. The first method requires two properties
to be satisfied. The first property is that two conservation laws
written in the form of differentials are required to be invariant
under these transformations. This property gives the representation
of the coefficients of the generator corresponding to the differentials.
Using these coefficients, the prolongation of the generator is obtained.
The second method provides a generalization of the first one, where
none of the assumptions about the differentials are required. This
method can also be applied to systems without knowing their conservation
laws. The solution obtained by the second method allowed us to state
the theorem about all reciprocal transformations for the two-dimensional
stationary gas dynamics equations.

The proposed methods provide systematic tools for finding reciprocal
transformations. The developed approach can be also extended to equations
with more than two independent variables.

\section*{Acknowledgment}

The authors are very thankful to Professor Colin Rogers for attracting
our attention to reciprocal transformations. The research was supported
by the Russian Science Foundation Grant No. 18-11-00238 `Hydrodynamics-type
equation: symmetries, conservation laws, invariant difference schemes'.

\pagebreak{}

\section*{Appendix }


Equations (\ref{eq:jul19.5}) can be written in the forms
\[
R_{\rho}=(R\,{}^{2}U\,{}^{2}(a_{35}p^{2}-2a_{34}p+2a_{33})+R\,{}^{2}V\,{}^{2}(a_{35}p^{2}-2a_{34}p+2a_{33}))/(2\rho^{2}(u^{2}+v^{2})),
\]
\[
R_{u}=(-R_{v}v+R\,{}^{2}U\,{}^{2}(-a_{35}p+a_{34})+R\,{}^{2}V\,{}^{2}(-a_{35}p+a_{34}))/u,
\]
\[
R_{p}=R\,{}^{2}a_{35}(U\,{}^{2}+V\,{}^{2})/2,
\]
\[
U_{\rho}=(PU(a_{35}p^{2}-2a_{34}p+2a_{33})+U(a_{45}p^{2}-2a_{44}p+2a_{43}))/(2\rho^{2}(u^{2}+v^{2})),
\]
\[
U_{u}=(-U_{v}v+PU(-a_{35}p+a_{34})+U(-a_{45}p+a_{44}))/u,
\]
\[
U_{p}=(PUa_{35}+Ua_{45})/2,
\]
\[
V_{\rho}=(PV(a_{35}p^{2}-2a_{34}p+2a_{33})+V(a_{45}p^{2}-2a_{44}p+2a_{43}))/(2\rho^{2}(u^{2}+v^{2})),
\]
\[
V_{u}=(-V_{v}v+PV(-a_{35}p+a_{34})+V(-a_{45}p+a_{44}))/u,
\]
\[
V_{p}=(PVa_{35}+Va_{45})/2,
\]
\[
P_{\rho}=(P\,{}^{2}(a_{35}p^{2}-2a_{34}p+2a_{33})+2P(a_{45}p^{2}-2a_{44}p+2a_{43})+2(a_{55}p^{2}-2a_{54}p+2a_{53}))/(2\rho^{2}(u^{2}+v^{2})),
\]
\[
P_{u}=(-P_{v}v+P\,{}^{2}(-a_{35}p+a_{34})+2P(-a_{45}p+a_{44})+2(-a_{55}p+a_{54}))/u,
\]
\[
P_{p}=(P\,{}^{2}a_{35}+2Pa_{45}+2a_{55})/2,
\]
\[
H_{\rho}=0,\,\,\,H_{u}=-H_{v}v/u,\,\,\,H_{p}=0,
\]
\[
\begin{array}{c}
^{x}f_{\rho}^{dx}=({}^{x}f^{dx}P(-a_{35}p^{2}+2a_{34}p-2a_{33})+{}^{x}f^{dx}RV\,{}^{2}(-a_{35}p^{2}+2a_{34}p-2a_{33})\\
+^{x}f^{dx}(-a_{45}p^{2}+2a_{44}p-2a_{43}+2\rho v^{2})-2{}^{y}f^{dx}\rho uv\\
+^{x}f^{dy}RUV(a_{35}p^{2}-2a_{34}p+2a_{33}))/(2\rho^{2}(u^{2}+v^{2})),
\end{array}
\]
\[
\begin{array}{c}
^{x}f_{u}^{dx}=(-{}^{x}f_{v}^{dx}v+{}^{x}f^{dx}P(a_{35}p-a_{34})+{}^{x}f^{dx}RV\,{}^{2}(a_{35}p-a_{34})+{}^{x}f^{dx}(a_{45}p-a_{44}+1)\\
+^{x}f^{dy}RUV(-a_{35}p+a_{34}))/u,
\end{array}
\]
\[
\begin{array}{c}
^{x}f_{p}^{dx}=(-{}^{x}f^{dx}Pa_{35}-{}^{x}f^{dx}RV\,{}^{2}a_{35}-{}^{x}f^{dx}a_{45}+{}^{x}f^{dy}RUVa_{35})/2,\end{array}
\]
\[
\begin{array}{c}
^{y}f_{\rho}^{dx}=(-2{}^{x}f^{dx}\rho uv+{}^{y}f^{dx}P(-a_{35}p^{2}+2a_{34}p-2a_{33})+{}^{y}f^{dx}RV\,{}^{2}(-a_{35}p^{2}+2a_{34}p-2a_{33})\\
+^{y}f^{dx}(-a_{45}p^{2}+2a_{44}p-2a_{43}+2\rho u^{2})+{}^{y}f^{dy}RUV(a_{35}p^{2}-2a_{34}p+2a_{33}))/(2\rho^{2}(u^{2}+v^{2})),
\end{array}
\]
\[
\begin{array}{c}
^{y}f_{u}^{dx}=(-{}^{y}f_{v}^{dx})v+{}^{y}f^{dx}P(a_{35}p-a_{34})+{}^{y}f^{dx}RV\,{}^{2}(a_{35}p-a_{34})+{}^{y}f^{dx}(a_{45}p-a_{44}+1)\\
+^{y}f^{dy}RUV(-a_{35}p+a_{34}))/u,
\end{array}
\]
\[
\begin{array}{c}
^{y}f_{p}^{dx}=(-{}^{y}f^{dx}Pa_{35}-{}^{y}f^{dx}RV\,{}^{2}a_{35}-{}^{y}f^{dx}a_{45}+{}^{y}f^{dy}RUVa_{35})/2,\end{array}
\]
\[
\begin{array}{c}
^{x}f_{\rho}^{dy}=({}^{x}f^{dx}RUV(a_{35}p^{2}-2a_{34}p+2a_{33})+{}^{x}f^{dy}P(-a_{35}p^{2}+2a_{34}p-2a_{33})\\
+^{x}f^{dy}RU\,{}^{2}(-a_{35}p^{2}+2a_{34}p-2a_{33})+{}^{x}f^{dy}(-a_{45}p^{2}+2a_{44}p-2a_{43}+2\rho v^{2})\\
-2^{y}f^{dy}\rho uv)/(2\rho^{2}(u^{2}+v^{2})),
\end{array}
\]
\[
\begin{array}{c}
^{x}f_{u}^{dy}=(-{}^{x}f_{v}^{dy}v+{}^{x}f^{dx}RUV(-a_{35}p+a_{34})+{}^{x}f^{dy}P(a_{35}p-a_{34})+{}^{x}f^{dy}RU\,{}^{2}(a_{35}p-a_{34})\\
+^{x}f^{dy}(a_{45}p-a_{44}+1))/u,
\end{array}
\]
\[
\begin{array}{c}
^{x}f_{p}^{dy}={}^{x}f_{p}^{dy}=({}^{x}f^{dx}RUVa_{35}-{}^{x}f^{dy}Pa_{35}-{}^{x}f^{dy}RU\,{}^{2}a_{35}-{}^{x}f^{dy}a_{45})/2,\end{array}
\]
\[
\begin{array}{c}
^{y}f_{\rho}^{dy}=({}^{y}f^{dx}RUV(a_{35}p^{2}-2a_{34}p+2a_{33})-2{}^{x}f^{dy}\rho uv+{}^{y}f^{dy}P(-a_{35}p^{2}+2a_{34}p-2a_{33}))\\
+^{y}f^{dy}RU\,{}^{2}(-a_{35}p^{2}+2a_{34}p-2a_{33})+{}^{y}f^{dy}(-a_{45}p^{2}+2a_{44}p-2a_{43}+2\rho u^{2}))/(2\rho^{2}(u^{2}+v^{2})),
\end{array}
\]
\[
\begin{array}{c}
^{y}f_{u}^{dy}=(-{}^{y}f_{v}^{dy}v+{}^{y}f^{dx}RUV(-a_{35}p+a_{34})+{}^{y}f^{dy}P(a_{35}p-a_{34})+{}^{y}f^{dy}RU\,{}^{2}(a_{35}p-a_{34})\\
+^{y}f^{dy}(a_{45}p-a_{44}+1))/u,
\end{array}
\]
\[
\begin{array}{c}
^{y}f_{p}^{dy}=({}^{y}f^{dx}RUVa_{35}-{}^{y}f^{dy}Pa_{35}-{}^{y}f^{dy}RU\,{}^{2}a_{35}-{}^{y}f^{dy}a_{45})/2,\end{array}
\]


Equations (\ref{eq:jul19.6}) are

\[
a_{44}a_{33}-a_{43}a_{34}-a_{33}=0,\,\,\,a_{54}a_{33}-a_{53}a_{34}-a_{43}=0,\,\,\,a_{54}a_{43}-a_{53}a_{44}-a_{53}=0,
\]
\[
a_{45}a_{33}-a_{43}a_{35}-a_{34}=0,\,\,\,a_{55}a_{33}-a_{53}a_{35}-a_{44}=0,\,\,\,a_{55}a_{43}-a_{54}-a_{53}a_{45}=0,
\]
\[
a_{45}a_{34}-a_{44}a_{35}-a_{35}=0,\,\,\,a_{55}a_{34}-a_{54}a_{35}-a_{45}=0,\,\,\,a_{55}a_{44}-a_{55}-a_{54}a_{45}=0.
\]


\end{document}